\documentclass[twocolumn]{autart} 

\usepackage{graphicx,color}      
\usepackage{cite,amssymb,amsmath}
\usepackage{algorithm}
\usepackage{algpseudocode}

\newtheorem{secthm}{Theorem}[section]
\newtheorem{seccor}[secthm]{Corollary}

\newtheorem{secprop}[secthm]{Proposition}
\newtheorem{secprob}[secthm]{Problem}
\newtheorem{secdefn}[secthm]{Definition}
\newtheorem{secrem}[secthm]{Remark}

\newcommand{\bA} { {\mathbb A}}
\newcommand{\bE} { {\mathbb E}}

\newcommand{\bP} { {\mathbb P}}
\newcommand{\bR} { {\mathbb R}}

\newcommand{\bZ} { {\mathbb Z}}

\newcommand{\cGP} { \mathcal{GP}}

\newcommand{\cN} { {\mathcal N}}

\def\red{\hfill $\lhd$}
\allowdisplaybreaks[4]

\begin{document}
\begin{frontmatter}

\title{An LMI Framework for Contraction-based Nonlinear Control Design by Derivatives of Gaussian Process Regression \thanksref{footnoteinfo}} 

\thanks[footnoteinfo]{This work was supported in part by JSPS KAKENHI Grant Numbers JP21H04875 and JP21K14185. }

\author[First]{Yu Kawano} 
\author[Second]{Kenji Kashima} 

\address[First]{Graduate School of Advanced Science and Engineering, Hiroshima University, Higashi-Hiroshima 739-8527, Japan (email: ykawano@hiroshima-u.ac.jp).}
\address[Second]{Graduate School of Informatics, Kyoto University, Kyoto 606-8501, Japan (e-mail: kk@i.kyoto-u.ac.jp).}

\begin{abstract}                
Contraction theory formulates the analysis of nonlinear systems in terms of Jacobian matrices. Although this provides the potential to develop a linear matrix inequality (LMI) framework for nonlinear control design, conditions are imposed not on controllers but on their partial derivatives, which makes control design challenging.  In this paper, we illustrate this so-called integrability problem can be solved by a non-standard use of Gaussian process regression (GPR) for parameterizing controllers and then establish an LMI framework of contraction-based control design for nonlinear discrete-time systems, as an easy-to-implement tool. Later on, we consider the case where the drift vector fields are unknown and employ GPR for functional fitting as its standard use. GPR describes learning errors in terms of probability, and thus we further discuss how to incorporate stochastic learning errors into the proposed LMI framework.
\end{abstract}


\end{frontmatter}
\section{Introduction}
Contraction theory~\cite{LS:98,FS:14}  has attracted massive research attention in the systems and control community, e.g., \cite{Bullo:22,TCS:21}, which establishes a differential geometric approach to study incremental properties, i.e., properties between any pair of trajectories. 
Revisiting nonlinear control theory from contraction perspectives brings new insights not only for stability analysis but also for dissipativity theory~\cite{Schaft:13,FS:18,KCS:21}, balancing theory~\cite{KS:17, Kawano:21}, and monotone systems~\cite{FS:15,KBC:20,KC:22} to name a few.
As an advantage in comparison with classical Lyapunov theory, the nature of incremental stability enables us to formulate various design problems in a unified framework, such as stabilizing control~\cite{MS:17,KO:17}, tracking control~\cite{GAA:21, RSJ:17}, observer design~\cite{AJP:16,LS:98}, and control design for achieving synchronizations \cite{AJP:16} and other rich behaviors \cite{FS:18}.
Owing to the differential geometric feature, these problems are described in terms of Jacobian matrices, which is expected as another advantage in practical use.
Indeed, restricting the class of controllers (and observer gains) into \emph{linear} reduces design problems to linear matrix inequalities (LMIs) \cite{Kawano:21,FS:18}.
However, \emph{nonlinear} control design is more involved because of the so-called integrability problem.
Namely, design conditions are imposed not on controllers but on their partial derivatives, which are the main difficulty in developing an LMI framework for contraction-based nonlinear control design. 

To overcome this difficulty, in this paper, we employ Gaussian process regression (GPR), a functional fitting tool \cite{RW:06, Bishop:06}.
As its non-standard use, we employ GPR to parametrize a controller based on its two important features: 1) computing derivatives of GPR is easy;
2) GPR becomes linear with respect to parameters while it possesses the flexibility to describe a nonlinear function.
Utilizing these two, we describe a condition for control design in terms of LMIs with respect to parameters of GPR.
Namely, we establish an LMI framework for contraction-based nonlinear stabilizing control design by explicitly addressing the integrability problem. 
We mainly consider nonlinear discrete-time systems with constant input vector fields and constant metrics for contraction.
Then, we mention that the formulations are still in the LMI framework for discrete-time systems with non-constant input vector fields
and continuous-time systems with non-constant input vector fields and non-constant metrics.
The proposed method is further applicable to the aforementioned various design problems thanks to a unified problem formulation by contraction theory.

In systems and control, GPR is typically used to estimate an unknown drift vector field from measured states, and 
\cite{IFT:17,TF:17,SKS:21,BKH:19,UH:19,UPH:18} study control design for learned models.
Other applications are a joint estimation of the state and model~\cite{Berntorp:21, BMT:20} and solving the Hamilton-Jacobi equations/inequalities \cite{IFY:18,FBT:18}. 
In particular, \cite{SKS:21,BKH:19,UH:19,UPH:18} study closed-loop stability under learning errors.
However, controllers are designed without taking learning errors into account, and learning errors are used for closed-loop analysis only.
Motivated by these works, in this paper, we also consider the case where a drift vector field is unknown, and this is learned by GPR.
In contrast to the conventional approach, we compensate for the learning error by control design,
which are benefits of developing the LMI framework for control design and learning models by GPR.
The proposed approach can be generalized to the case where the whole system dynamics are unknown, and only system's input and output are measurable, since
there are learning approaches by GPR in such a setting \cite{FLS:14,FCR:14}.

As relevant researches, \cite{TC:19,TCS:20,SKS:21} give neural network frameworks for contraction-based control design, 
which requires iterations for finding suitable parameters in contrast to the proposed LMI framework. 
The paper \cite{IFT:17} formulates control design for models learned by GPR in an LMI framework and designs a switching linear controller, but  does not use GPR for control design or is not based on contraction theory.
Therefore, the result in \cite{IFT:17} is not applicable for solving the integrability problem.
In this paper, we establish an LMI framework for contraction-based control design by utilizing the derivatives of GPR to solve the integrability problem.

The remainder of this paper is organized as follows. In Section~\ref{PR:sec}, we pose the problem formulation by mentioning the integrability problem of control design in contraction theory. In Section~\ref{GP:sec}, we develop an LMI framework for contraction-based control design by utilizing derivatives of GPR.
In Section~\ref{CDUS:sec}, we consider the case where drift vector fields are unknown. 
In Section~\ref{EX:sec}, the proposed control design method is illustrated by the means of an example.

{\it Notation:}
The sets of real numbers, non-negative integers, and positive integers are denoted by~$\bR$, $\bZ_{\ge 0}$, and~$\bZ_{> 0}$, respectively. The identity matrix with the size $n$ is denoted by~$I_n$. For~$P, Q \in \bR^{n \times n}$,  $P \succ Q$ (resp. $P \succeq Q$) means that $P - Q$ is symmetric and positive definite (resp. semi-definite). The Euclidean norm of $x \in \bR^n$ weighted by $P \succ 0$ is denoted by~$|x|_P:=\sqrt{x^\top P x}$. If $P = I_n$, this is simply denoted by $|x|$. 
The Moore-Penrose inverse of a matrix $A$ is denoted by $A^+$.

For a function $f(x,y)$, the row vector-valued function consisting of its partial derivatives with respect to $x$ is denoted by 
$\partial_x f:= \partial f/\partial x = [\begin{matrix} \partial f/\partial x_1 & \cdots & \partial f/\partial x_n \end{matrix}]$. 
If $f$ depends on $x$ only, this is simply denoted by $\partial f$.
For a scalar-valued function $k(x, x')$, its Hessian matrix is denoted by 
$\partial^2 k(x, x') := \partial^2 k(x, x')/\partial x \partial x'$, which is a matrix-valued function.
The multivariate normal distribution with mean $\mu$ and variance $\Sigma$ is denoted by $\cN (\mu, \Sigma)$.  The (standard) expectation is denoted by~$\bE [\cdot ]$. 
A stochastic process~$\{\omega_k\}_{k \in \bZ_{\ge 0}}$ is said to be i.i.d. if $\omega_k$, $k \in \bZ_{\ge 0}$ are independently distributed, and none of the characteristics of $\omega_k$ changes with $k \in \bZ_{\ge 0}$.

\section{Preliminaries}\label{PR:sec}
\subsection{Problem Formulation}\label{PF:sec}
Consider the following discrete-time nonlinear system:
\begin{align}\label{sys}
x_{k+1} = f (x_k) + b u_k,
\quad k \in \bZ_{\ge 0},
\end{align}
where $x_k \in \bR^n$ and $u_k \in \bR$ denote the state and input, respectively;
$f:\bR^n \to \bR^n$ is of class $C^1$, i.e., continuously differentiable, and $b \in \bR^n$. 
The $i$th components of $f$ and $b$ are denoted by $f_i$ and $b_i$, respectively.
For the sake of notational simplicity, we consider single-input systems.
However, the results can readily be generalized to multiple-input systems as explained below.
We also later discuss the case where $b$ is a function of $x$.

In contraction theory \cite{LS:98,TRK:18,KH:21}, we study incremental stability as a property of the pair of trajectories, stated below.
\begin{secdefn}
The system $x_{k+1} = f(x_k)$ (with its copy $x'_{k+1} = f(x'_k)$) is said to be \emph{incrementally exponentially stable} (IES) if there exist $a>0$ and $\lambda \in (0,1)$ such that
\begin{align*}
|x_k-x'_k| \le a \lambda^k |x_0 - x'_0|, 
\quad \forall k \in \bZ_{\ge 0}
\end{align*}
for each~$(x_0, x'_0) \in \bR^n \times \bR^n$.
\red
\end{secdefn}

Applying \cite[Theorem 15]{TRK:18} to control design yields the following IES condition.
\begin{secprop}\label{IES:prop}
Suppose that there exist $\varepsilon > 0$, $P \in \bR^{n \times n}$,
and $p_\partial: \bR^n \to \bR^{1 \times n}$ of continuous such that
\begin{align}\label{IES:cond}
&\begin{bmatrix}
P & * \\
\partial f (x) P  + b  p_\partial (x) & P
\end{bmatrix}
\succeq \varepsilon I_{2n}
\end{align}
for all $x \in \bR^n$, where $*$ represents the appropriate matrix.
If there exists $p: \bR^n \to \bR$ of class $C^1$ such that
\begin{align}\label{integrability}
p_\partial (x) = \partial p (x) P
\end{align}
for all $x \in \bR^n$, 
then the closed-loop system $x_{k+1} = f(x_k) + b p(x_k)$ is IES.
\end{secprop}

\begin{pf}
By the Schur complement, the set of \eqref{IES:cond} and \eqref{integrability} is equivalent to $P \succ 0$ and
\begin{align}\label{IES:cond2}
&P (\partial f (x)  + b \partial p(x) )^\top P^{-1}
(\partial f (x)  + b \partial p(x) ) P\nonumber\\
&- P \preceq  \varepsilon I_n.
\end{align}
This is nothing but the definition of uniform contraction with $\Theta = P^{-1/2}$ \cite[Definition 6]{TRK:18} 
for the closed-loop system $x_{k+1} = f(x_k) + b p(x_k)$.
Therefore, \cite[Theorem 15]{TRK:18} concludes IES of the closed-loop system.
\qed
\end{pf}

In \cite[Theorem 15]{TRK:18},  it has been shown that a closed IES system admits 
a state-dependent $P$ satisfying a similar inequality as \eqref{IES:cond}.
Moreover, such a $P$ is uniformly lower and upper bounded by constant matrices.
It is not yet clear when $P$ becomes constant. 
If one restricts the class of controllers into linear for a constant $P$, 
control design can be reduced to linear matrix inequalities (LMIs); see, e.g., \cite{FS:18,Kawano:21}.
Namely, control design can be sometimes formulated as a practically solvable problem even for nonlinear systems.
Indeed, \eqref{IES:cond} is an LMI with respect to $\varepsilon$, $P$ and $p_\partial (x)$ at each $x \in \bR^n$. 
However, as in Proposition \ref{IES:prop}, 
designing nonlinear controllers is not fully formulated in the LMI framework, 
due to the so-called integrability constraint \eqref{integrability}, i.e., 
$p_\partial (x) P^{-1}$ needs to be the partial derivative of some function $p(x)$ providing a feedback control law $u = p(x)$. 
The main objective of this paper is to develop an LMI framework for 
stabilizing nonlinear control design by tackling the integrability constraint, stated below.

\begin{secprob}\label{main:prob}
Given $f(x)$ and $b$ of the system \eqref{sys}, develop an LMI framework for designing 
$p:\bR^n \to \bR$ which satisfies all the conditions in Proposition~\ref{IES:prop},
if such $p$ exists. \red
\end{secprob}

\begin{secrem}
We later consider the case where $f$ is unknown by learning it.
As a byproduct of developing control design methodologies in the LMI framework,
it is possible to compensate for learning errors by control design.
\red
\end{secrem}

An important feature of contraction theory is to study the convergence between any pair of trajectories.
By virtue of this, one can handle observer design \cite{AJP:16,LS:98}, tracking control \cite{GAA:21,RSJ:17}, and control design for imposing synchronizations \cite{AJP:16}
and rich behavior such as limit cycles \cite{FS:18} in the same framework as stabilizing control design.
These references mainly focus on continuous-time systems, but similar results can be delivered to discrete-time systems, since incremental stability conditions in contraction analysis have been derived also for discrete-time systems \cite{LS:98,TRK:18}.
Moreover, the proposed method in this paper can be generalized to continuous-time systems as will be explained in Section~\ref{ss:CD}. 
Therefore, solving Problem~\ref{main:prob} can result in LMI frameworks for various design problems.

Integrability constraints sometimes appear in nonlinear adaptive control or observer design.
Since directly addressing integrability constraints are difficult, 
there are techniques for avoiding them by adding the dynamic order of the identifier. 
However, as pointed out by \cite{Krstic:09}, adding additional dynamics can degenerate control performances,
and it has not been validated that such an approach works for contraction-based control design.
Therefore, it is worth solving Problem~\ref{main:prob} directly.
In particular, we provide an LMI framework for control design, which is easy-to-implement.
The proposed method may be tailored for nonlinear adaptive control or observer design 
although this is beyond the scope of this paper.

\subsection{Gaussian Process Regression}
To solve Problem~\ref{main:prob}, we employ Gaussian process regression (GPR) \cite{RW:06, Bishop:06}.
We, in this subsection, briefly summarize basics of GPR and, in the next section, demonstrate that 
GPR is a suitable tool for handling problems involving partial derivatives, 
e.g., integrability conditions.
 
Let $\{x^{(i)} \}_{i=1}^N$, $x^{(i)} \in \bR^n$ be input data, and 
let $\{y^{(i)} \}_{i=1}^N$, $y^{(i)} \in \bR$ be the corresponding output data given by
\begin{align}\label{y}
y^{(i)} = p (x^{(i)}) +  \omega^{(i)} , 
\quad  i=1, \dots, N,
\end{align}
where $\omega^{(i)} \sim \cN (0, \sigma^2 )$ is i.i.d.
GPR is a technique to learn an unknown function $p:\bR^n \to \bR$ from input-output data $\{x^{(i)},  y^{(i)}\}_{i=1}^N$
by assuming $p$ as a \emph{Gaussian process} (GP).

A stochastic process $p$ is said to be GP if 
any finite set $\{p(x^{(i)})\}_{i=1}^N$ has a joint Gaussian distribution~\cite[Definition 2.1]{RW:06}.
A GP is completely specified by its mean function $m_p: \bR^n \to \bR$ and covariance function $k_p: \bR^n \times \bR^n \to \bR$.
They are defined by
\begin{align}
m_p(x) &= \bE [p(x)]
\label{p:pmean}\\
k_p(x, x')&= \bE [(p(x) - m_p(x)) (p(x') - m_p(x'))],
\label{p:variance}
\end{align}
and we represent the GP by $p \sim \cGP (m_p(x), k_p(x,x'))$.

The essence of GPR is to estimate $m_p(x)$ and $k_p(x,x')$ as 
the posterior mean and covariance given~$\{x^{(i)},  y^{(i)}\}_{i=1}^N$ by the Bayes estimation \cite{RW:06, Bishop:06}.
Typically, the prior mean $m_0(x)$ is selected as zero.
The prior covariance $k_0 (x,x')$ needs to be a positive definite kernel  \cite[Section 6]{RW:06}, and we also require smoothness.
Standard kernels are linear, polynomial, or squared exponential (SE) \cite{RW:06, Bishop:06}, which are all smooth and positive definite.
For instance, an SE kernel is
\begin{align}\label{SE}
k_0 (x, x') = \beta e^{- | x - x'|_{\Sigma^{-1}}^2/2},
\end{align}
where $\beta >0$ and $\Sigma \succ 0$ in addition to $\sigma >0$ in \eqref{y} are free parameters, called hyper parameters.
The hyper parameters can be selected to maximize the marginal likelihood from observed data; see e.g., \cite[Section 5.4]{RW:06}.

To use GPR for learning $p(x)$, we need its data as in \eqref{y}. 
However, Proposition~\ref{IES:prop} contains no information of $p(x)$ but of $\partial p(x)$ via 
the integrability constraint \eqref{integrability}.
To handle this situation, we employ derivatives of GPR as elaborated in the next section.

\section{Contraction-based Nonlinear Control Design}\label{GP:sec}
In this section, we establish an LMI framework for contraction-based nonlinear control design by
utilizing derivatives of GPR to solve the integrability problem.
Then, we discuss generalizations of the proposed method to non-constant input vector fields and continuous-time cases.

\subsection{LMI Frameworks for Nonlinear Control Design}
Taking the partial derivatives of \eqref{p:pmean} and \eqref{p:variance}, 
we have the joint distribution of $p(x)$ and $\partial p(x)$ as in
\begin{align}\label{joint}
\begin{bmatrix}
p \\ \partial^\top p
\end{bmatrix}
\sim \cGP \left(
\begin{bmatrix}
m_p (x) \\ \partial^\top m_p(x)
\end{bmatrix}, 
\begin{bmatrix}
k_p (x,x') & \partial_{x'} k_p (x,x') \\
\partial_x^\top k_p (x,x') & \partial^2 k_p (x,x')
\end{bmatrix}\right),
\end{align}
see, e.g., \cite[Equation (2)]{PZH:21}.
The goal of this subsection is to find $m_p(x)$ based on the Bayes estimation such that 
$p(x) = m_p(x)$ is a solution to Problem~\ref{main:prob}.
For the sake of notational simplicity, we select the prior mean of $p(x)$ as zero.

A standard procedure of the Bayes estimation is that we first select a class $C^2$ positive definite kernel $k_0(x,x')$ as a prior covariance. 
Then, we compute the posterior mean $m_p(x)$ given data of $\partial p(x)$. 
Looking at this from a different angle, $m_p(x)$ can be viewed as a function of data of $\partial p(x)$.
Based on this perspective, we consider generating suitable data such that $m_p(x)$ becomes a solution to Problem~\ref{main:prob}
as a non-standard use of GPR.

To this end, let $\{x^{(i)}, \hat p^{(i)} \}_{i=1}^N$ denote a data set to be generated, where
\begin{align}\label{yp}
\hat p^{(i)} = \partial p (x^{(i)}) + \omega_p^{(i)}, 
\quad i=1, \dots, N.
\end{align}
The role of i.i.d. $\omega_p^{(i)} \sim \cN (0, \sigma_p^2 I_n)$ is explained later; 
one can simply choose it as zero.
Define the vector consisting of $\{\hat p^{(i)}\}_{i=1}^N$ by
\begin{align*}
Y_p:= 
\begin{bmatrix}
\hat p^{(1)}  & \cdots & \hat p^{(N)}
\end{bmatrix} \in \bR^{nN}.
\end{align*}
Also, we introduce the following notations:
\begin{align}
&X_N:=
\begin{bmatrix}
(x^{(1)})^\top  & \cdots & (x^{(N)})^\top
\end{bmatrix}^\top
\in \bR^{nN} \nonumber\\
&\partial p(X_N):=
\begin{bmatrix}
\partial p (x^{(1)}) & \cdots & \partial p (x^{(N)})
\end{bmatrix}
\in \bR^{nN} \nonumber\\
&{\bf k}'_{\partial, 0} (x'):= 
\begin{bmatrix}
\partial_x k_0 (x^{(1)},  x') & \cdots & \partial_x k_0 (x^{(N)}, x')
\end{bmatrix} \nonumber\\
&{\bf k}_{\partial, 0} (x):= 
\begin{bmatrix}
\partial_{x'} k_0 (x,  x^{(1)}) & \cdots & \partial_{x'} k_0 (x, x^{(N)})
\end{bmatrix} \nonumber\\
&K_0 := 
\begin{bmatrix}
\partial^2 k_0 (x^{(1)}, x^{(1)}) & \cdots & \partial^2 k_0 ( x^{(1)}, x^{(N)})\\
\vdots &  & \vdots \\
\partial^2 k_0 (x^{(N)}, x^{(1)}) & \cdots & \partial^2 k_0 ( x^{(N)}, x^{(N)})
\end{bmatrix}\nonumber\\
&\hspace{50mm} \in \bR^{n N \times n N}.
\label{K}
\end{align}
Then, the joint distribution of the prior distribution of $p(x)$, denoted by $p^{\rm pri}(x)$ and $\partial p(X_N)$ is
\begin{align*}
\begin{bmatrix}
p^{\rm pri} \\ \partial^\top p(X_N)
\end{bmatrix}
\sim \cN \left(
\begin{bmatrix}
0 \\ Y_p^\top
\end{bmatrix}, 
\begin{bmatrix}
k_0 (x,x') & {\bf k}_{\partial, 0} (x) \\
({\bf k}'_{\partial, 0})^\top (x') & K_0 + \sigma_p^2 I_{nN}
\end{bmatrix}\right).
\end{align*}
By the Bayes estimation, we can compute the posterior mean $m_p(x)$ given $Y_p$ as
\begin{align}\label{p:mean}
m_p (x | Y_p) := \sum_{i=1}^N \partial_{x'} k_0 (x, x^{(i)}) h_p^{(i)},
\end{align}
where $h_p^{(i)}$ denotes the $i$th block-component with size $n$ of $h_p$ defined by
\begin{align}
h_p := (K_0 + \sigma_p^2 I_{nN} )^{-1} Y_p^\top \in \bR^{nN}.
\label{h}
\end{align}
The partial derivative of $m_p (x | Y_p)$ is easy-to-compute:
\begin{align}\label{dp:mean}
\partial m_p (x | Y_p) = \frac{\partial }{\partial x} \left( \sum_{i=1}^N \partial_{x'} k_0 (x, x^{(i)}) h_p^{(i)} \right). 
\end{align}
Especially if $\sigma_p = 0$ and $K_0 \succ 0$, it follows from~\eqref{dp:mean} that
\begin{align}\label{dp:mean2}
\partial m_p (x^{(i)} | Y_p) = \hat p^{(i)},
\quad i = 1, \dots, N. 
\end{align}
Note that $m_p(x|Y_p)$ and $\partial m_p(x |Y_p)$ are nonlinear functions of $x$ and linear functions of $Y_p$. Therefore, control design reduces to generating suitable $Y_p$, i.e., $\{x^{(i)}, \hat p^{(i)} \}_{i=1}^N$, which is proceeded based on Proposition \ref{IES:prop}.

Now, we are ready to develop an LMI framework for contraction-based nonlinear control design.
Substituting $p_\partial = \partial m_p (x|Y_p) P$ into \eqref{IES:cond} does not give
an LMI because of the coupling between $Y_p$ and $P$.
This issue can be addressed  by using a standard technique of an LMI.
According to \cite[Theorem 2.3.11]{RIK:17}, \eqref{IES:cond}, i.e., the set of $P \succ 0$ and \eqref{IES:cond2} implies, 
for all $x \in \bR^n$, 
\begin{align}\label{forP}
\left\{\begin{array}{l}
P \succ 0 \\
b^+ (P - \partial f(x) P \partial^\top f(x)) (b^+)^\top \succeq  \varepsilon_p I_{n-1}.
\end{array}\right.
\end{align}
This is an LMI with respect to $P$ and $\varepsilon_p>0$ at each $x \in \bR^n$.
For a solution $P$ to \eqref{forP} and $p_\partial = \partial m_p (x| Y_p) P$, 
one only has to solve \eqref{IES:cond} with respect to $Y_p$.
The proposed procedure for solving Problem~\ref{main:prob} is summarized in Algorithm~\ref{alg:main}.

\begin{algorithm}[htbp]
  \caption{Solving Problem \ref{main:prob}}
  \label{alg:main}
  \begin{algorithmic}[1]
    \Require{$f(x)$, $b$, $k_0(x, x')$, $\sigma_p \ge 0$, $\{x^{(i)}\}_{i=1}^N$}
    \Ensure{$p(x)$ and $P$ (if they exist)}
       \State {Define $\partial m_p (x|Y_p)$ as in \eqref{dp:mean}}
        \State{Solve a finite family of LMIs with respect to $P$ and $\varepsilon_p>0$:
        \begin{align}\label{LMI1}
        &\left\{\begin{array}{l}
        P \succ 0 \\
        b^+ (P - \partial f(x^{(i)}) P \partial^\top f(x^{(i)})) (b^+)^\top \succeq \varepsilon_p I_{n-1}\\
        \end{array}\right. \nonumber\\
        &\hspace{45mm} i = 1, \dots, N
        \end{align}}
        \State {For $P$ obtained in Step 2, solve a finite family of LMIs with respect to $Y_p$ and $\varepsilon >0$:
        \begin{align}\label{LMI2}
        \begin{bmatrix}
        P & * \\
         ( \partial f (x^{(i)})  + b \partial m_p (x^{(i)}| Y_p)  )P & P
        \end{bmatrix}
        \succeq \varepsilon I_{2n},\\
        i = 1, \dots, N
        \nonumber
        \end{align}}
          \State {Define $m_p (x| Y_p)$ as in \eqref{p:mean} by using $Y_p$ obtained in Step 3}
    \State Return $p(x):=m_p(x| Y_p)$ and $P$
  \end{algorithmic}
\end{algorithm}

If Algorithm \ref{alg:main} has a set of solutions, the obtained $p(x)$ is a solution to Problem \ref{main:prob}
at each data point, stated below.
\begin{secthm}\label{main1:thm}
Given a system~\eqref{sys}, a class $C^2$ positive definite kernel $k_0(x, x')$, $\sigma_p \ge 0$, and $\{x^{(i)}\}_{i=1}^N$, 
suppose that Algorithm~\ref{alg:main} has a set of solutions $p(x)$ and $P$.
Then, $p(x)$ satisfies 
\eqref{IES:cond} and \eqref{integrability} for all $x= x^{(i)}$, $i=1,\dots,N$. 
\red
\end{secthm}

\begin{pf}
Using a set of solutions $p(x)$ and $P$, define $p_\partial (x) = \partial p(x) P$.
Then, \eqref{integrability} holds for all $x \in \bR^n$.
It suffices to confirm that~\eqref{IES:cond} holds at $x = x^{(i)}$, $i=1,\dots,N$.
Since $p(x) = m_p(x|Y_p)$, we have
$\partial p(x^{(i)}) = \partial m_p(x^{(i)}|Y_p)$, $i=1,\dots,N$.
Therefore, \eqref{LMI2} implies~\eqref{IES:cond} for $x = x^{(i)}$, $i=1,\dots,N$.
\qed
\end{pf}

Note that in Algorithm~\ref{alg:main}, the number $N$ of training data and data point set $\{x^{(i)}\}_{i=1}^N$ can be chosen arbitrarily.
As $N$ increases, the number of $Y_p \in \bR^{nN}$ in~\eqref{LMI2} increase. 
However, the problem is still convex.  
As explained in the next subsection, for sufficiently large $N$, 
one can show that $p(x) = m_p(x| Y_p)$ is a solution to Problem \ref{main:prob} other than $\{x^{(i)}\}_{i=1}^N$ when $x^{(i)}$ is distributed evenly in the state space.
Moreover, such $N$ can be found.

When $\sigma_p =0$ and $K_0 \succ 0$, \eqref{dp:mean2} helps to simplify Algorithm \ref{alg:main}.
In fact, we only have to solve a finite family of LMIs once, stated below without the proof.

\begin{seccor}\label{main1:cor}
Given a class $C^2$ positive definite kernel $k_0(x, x')$ and $\{x^{(i)}\}_{i=1}^N$, suppose that $K_0 \succ 0$, and 
the following finite family of LMIs admits
a set of solutions $\{\bar p^{(i)}\}_{i=1}^N$, $P \in \bR^{n \times n}$, and $\varepsilon > 0$:
\begin{align}\label{LMI3}
        \begin{bmatrix}
        P & * \\
        \partial f (x^{(i)}) P  + b \bar p^{(i)} & P
        \end{bmatrix}
        \succeq \varepsilon I_{2n}, 
        \quad
        i = 1, \dots, N.
\end{align}
Define $\hat p^{(i)}=\bar p^{(i)} P^{-1}$, $i=1,\dots,N$ and choose $\sigma_p=0$.
Then,  $p(x) := m_p(x| Y_p)$ defined by \eqref{p:mean} satisfies 
\eqref{IES:cond} and \eqref{integrability} for all $x= x^{(i)}$, $i=1,\dots,N$. 
\red
\end{seccor}

\begin{secrem}\label{p:rem}
It is not guaranteed that the constructed controller $p(x)$ in Theorem~\ref{main1:thm} preserves an equilibrium point $x^*$ of $x_{k+1} = f(x_k)$.
However, it is easy to impose $p(x^*) = 0$.
An approach is to use shifted $u = p(x) - p(x^*)$.
Another approach is to utilize \eqref{y}.
Consider new data $\{x^{(i)}, y^{(i)} \}_{i=N+1}^M$ in \eqref{y}.
Define $Y:=[\begin{matrix} y^{(N+1)} & \cdots & y^{(M)} \end{matrix}]^\top$.
Then, the posterior mean $m_p(x)$ given $Y$ and $Y_p$ can be computed by the Bayes estimation also.
As a special case, specifying $M=N+1$, $x^{(N+1)} = x^*$, $y^{(N+1)} = 0$, and $\sigma = 0$ can result $p(x^*) = 0$ for learned $p(x)$.
More generally, one can specify the values of $p(x)$ at arbitrary finite points  $\{x^{(i)} \}_{i=N+1}^M$.
\red
\end{secrem}

\begin{secrem}\label{mi:rem}
In the multiple-input case, each component $p_i$ of $p$ can be designed separately by introducing $m_{p,i}(x|Y_p)$
corresponding to $p_i$.
Namely, the results obtained in this paper can readily be generalized to the multiple-input case. 
\red
\end{secrem}

At the end of this subsection, we argue the role of $\omega_p^{(i)} \sim \cN (0, \sigma_p^2 I_n)$ in \eqref{yp}
by interpreting the procedure of Algorithm \ref{alg:main} as follows.
We generate suitable $Y_p$ based on the LMIs and then construct $p(x) = m_p(x|Y_p)$ by functional fitting.
In functional fitting, overfitting is a common issue, since a constructed function becomes unnecessarily complex.
The variance $\sigma_p^2$ specifies how much a constructed function needs to fit to data, 
and thus adding the noise $\omega_p^{(i)}$ helps to avoid overfitting. 
However, if $\sigma_p^2$ is large, a constructed function can fit to $\omega_p^{(i)}$ instead of $\partial p( x^{(i)})$. 
This is well known as the bias-variance tradeoff in functional fitting \cite[Section 5.4.4]{GDC:16}.
It is worth emphasizing that Theorem~\ref{main1:thm} holds for arbitrary $\sigma_p>0$.

If $K_0$ in \eqref{K} is non-singular, $h_p$ in \eqref{h} is nothing but the minimizer of the following optimization problem:
\begin{align}
\min_{h_p} |Y_p - K_0 h_p|_{K_0^{-1}}^2  + \sigma_p^2 |h_p|^2.
\label{opt}
\end{align}
The first term evaluates the fitting error at each $x^{(i)}$ and the second one is for regularization. Especially when $\sigma_p=0$, the optimal value becomes zero for the obtained $h_p$, i.e., a complete fitting $\partial p( x^{(i)}) = \hat p^{(i)}$, $i=1,\dots, N$ is achieved as stated by Corollary~\ref{main1:cor}. A reproducing kernel Hilbert space is a formal tool to study a functional fitting problem under regularization as an optimization problem. For the prior variance as a kernel, the pair $(p, \partial p)$ can be understood as a minimizer according to the representer theorem \cite[Theorem 2]{PDC:14}.

\subsection{Closed-loop Stability}\label{ss:CS}
In Theorem \ref{main1:thm}, it is not clear whether $p_\partial (x) = \partial p(x) P$ satisfies 
\eqref{IES:cond} other than $\{ x^{(i)} \}_{i=1}^N$ in contrast to the integrability condition \eqref{integrability}.
Applying standard arguments of the polytope approach \cite{BEF:94}, 
we improve the LMIs in Algorithm~\ref{alg:main} such that its solution satisfies \eqref{IES:cond} other than $\{ x^{(i)} \}_{i=1}^N$ 
when $N$ is sufficiently large.

For a set of matrices $\bA:= \{A_1,\dots,A_L \}$, let ${\rm ConvexHull} (\bA)$ denote its convex hull, i.e., 
\begin{align*}
&{\rm ConvexHull} (\bA) \\
&:= \left\{A = \sum_{\ell =1}^L \theta_\ell A_\ell:  \sum_{\ell =1}^L \theta_\ell = 1, \; \theta_\ell \ge 0, \ell=1,\dots,L \right\}.
\end{align*}
Now, we choose $L = N$ and 
\begin{align*}
A_i :=  \partial f ( x^{(i)}) + b \partial p ( x^{(i)}), 
\quad i=1, \dots, N.
\end{align*}
From the standard discussion of the polytope approach~\cite{BEF:94},
\eqref{IES:cond} and \eqref{integrability} hold for all $x \in \bR^n$ belonging to
\begin{align*}
D:= \{x \in \bR^n : \partial f (x) + b  \partial p(x) \in {\rm ConvexHull} (\bA) \}.
\end{align*}
This set $D$ can be made larger by increasing the number $N$ of data used for control design. 

Increasing $N$ is not the only approach to enlarge a set of $x$ in which \eqref{IES:cond} holds. 
Another approach is to improve the LMIs in Algorithm~\ref{alg:main}. 
First, we decide $\bA^{(i)} := \{A^{(i)}_1,\dots,A^{(i)}_{L^{(i)}}\} \subset \bR^{n \times n}$ which is arbitrary as long as
\begin{align}\label{relax}
\partial f (x^{(i)}) \in {\rm ConvexHull} (\bA^{(i)}), 
\quad i=1, \dots, N.
\end{align}
Instead of \eqref{LMI1}, we consider the following finite family of LMIs with respect to $P \in \bR^{n \times n}$ and $\varepsilon_p > 0$:
 \begin{align}\label{LMI1_cvx}
        \left\{\begin{array}{l}
        P \succ 0 \\
        b^+ (P - A^{(i)}_{\ell^{(i)}} P (A^{(i)}_{\ell^{(i)}})^\top ) (b^+)^\top \succeq \varepsilon_p I_{n-1} \\
        \end{array}\right. \\
        \ell^{(i)} = 1,\dots, L^{(i)}, \; i = 1, \dots, N.
        \nonumber
        \end{align}
Using a solution $P$, we consider the following finite family of LMIs with respect to $Y_p \in \bR^{nN}$ and $\varepsilon > 0$ instead of \eqref{LMI2}:
\begin{align}\label{LMI2_cvx}
\begin{bmatrix}
P &*  \\
( A^{(i)}_{\ell^{(i)}} + b \partial m_p (x^{(i)}| Y_p) ) P & P
\end{bmatrix}
\succeq \varepsilon I_{2n}\\
\quad 
\ell^{(i)} = 1,\dots, L^{(i)}, \; i = 1,\dots, N.
\nonumber
\end{align}
The newly obtained $p(x):=m_p (x| Y_p)$ satisfies \eqref{IES:cond} for all $x \in \bR^n$ belonging to
\begin{align*}
\bar D:= \bigcup_{i=1}^N \{x \in \bR^n : &\; \partial f (x) + b \partial p (x) \in {\rm ConvexHull} (\bA^{(i)}) \}.
\end{align*}
Now, we are ready to state a control design procedure such that the IES conditions \eqref{IES:cond} and \eqref{integrability} hold on a given bounded set.

\begin{secthm}\label{cvx:thm}
Consider a system~\eqref{sys}, a class $C^2$ positive definite kernel $k_0(x, x')$, and $\sigma_p \ge 0$.
Let $D \subset \bR^n$ denote an $n$-dimensional closed hypercube. 
We partition it evenly to $N$ closed hypercubes, denoted by $\{D^{(i)}\}_{i=1}^N$, where 
$N = r^n$, $r \in \bZ_{>0}$. For all $r \in \bZ_{>0}$, there exist $\bA^{(i)}$, $i=1,\dots,N$ such that
\begin{align}\label{relax2}
&\partial f (x) \in {\rm ConvexHull} (\bA^{(i)}) \\
&\hspace{10mm}\forall x \in D^{(i)}, \; 
\forall i=1, \dots, N. \nonumber
\end{align}
Let each $x^{(i)}$, $i=1, \dots, N$ be the center of $D^{(i)}$. 
Suppose that for such $\bA^{(i)}$, $i=1,\dots,N$, 
\begin{enumerate}
\renewcommand{\labelenumi}{\arabic{enumi})}
\item for all $r>0$, the LMI \eqref{LMI1_cvx} admits $P \in \bR^{n \times n}$ and $\varepsilon_p > 0$;

\item for all $r>0$ and $P$ obtained in item 1), the LMI~\eqref{LMI2_cvx} admits $Y_p \in \bR^{nN}$ and $\varepsilon > 0$;

\item for the elements $\hat p^{(i)}$ of $Y_p$, there exists a strictly decreasing positive function $\delta$ of $r$ such that $\delta (r) \to 0$ as $r \to \infty$, and $|\hat p^{(i)} - \hat p^{(j)}| < \delta (r)$ if $D^{(i)}$ and $D^{(j)}$ are next to each other.
\end{enumerate}
Then, there exists a sufficiently large $r > 0$ such that 
$p(x) := m_p(x| Y_p)$ satisfies \eqref{IES:cond} and \eqref{integrability} for all $x \in D$.
\end{secthm}

\begin{pf}
First, we show \eqref{relax2}. Since a continuous function is bounded on a bounded set, $\{\partial f (x) \in \bR^{n \times n}: x \in D^{(i)} \}$ is bounded for each $i=1,\dots, N$. Each bounded set admits its convex hull. Therefore, there exists $\bA^{(i)}$ satisfying \eqref{relax2}.

Next, we consider the latter statement. 
We choose $p_\partial (x) = \partial p(x) P= \partial m_p(x| Y_p) P$.
Then, \eqref{integrability} holds on $\bR^n$. 
It remains to show that~\eqref{IES:cond} holds on $D$.
Items 1) and 2) and \eqref{relax2} imply that for each $i=1,\dots,N$, 
\begin{align*}
\begin{bmatrix}
P & * \\
 \partial f (x) P  + b p_\partial (x) & P
\end{bmatrix}
\succeq \varepsilon I_{2n}, 
\quad \forall x \in D^{(i)}.
\end{align*}
The strict inequality holds for any positive $\bar \varepsilon < \varepsilon$. Since $p_\partial (x) = \partial m_p(x| Y_p) P$ is continuous (with respect to $x$), there exists a sufficiently small $\bar D_{\bar \varepsilon}^{(i)} \subset \bR^n$ centered at $x^{(i)}$ such that
\begin{align*}
\begin{bmatrix}
P & * \\
 \partial f (x) P +  b p_\partial (x) & P
\end{bmatrix}
\succ \bar \varepsilon I_{2n}, 
\quad \forall x \in \bar D_{\bar \varepsilon}^{(i)}.
\end{align*}
From item 3), the continuity of $\partial f$ and $p_\partial$, and the boundedness of $D^{(i)}$, there exists a sufficiently large $r \in \bZ_{> 0}$ such that $D^{(i)} \subset \bar D_{\bar \varepsilon}^{(i)}$ for all $i = 1, \dots, N$. Consequently, \eqref{IES:cond} hold on $D$, a union of $D^{(i)}$, $i=1,\dots,N$.
\qed
\end{pf}

In the above theorem, $D$ and $D^{(i)}$ are not necessarily to be hypercubes or closed. Essential requirements are that $D$ is covered by $\{D^{(i)}\}_{i=1}^N$, and each $D^{(i)}$ shrinks as $N$ increases. In the proof, we show that a switching controller $u = \partial p(x^{(i)}) x$, $x \in D^{(i)}$ also satisfies \eqref{IES:cond} and \eqref{integrability}. However, a continuous controller $u=p(x)$ is more easy-to-implement.

\begin{secrem}\label{local:rem}
In Theorem~\ref{cvx:thm}, IES of the closed-loop system is guaranteed on geodesically convex $D$ if 
either $D$ is positively invariant or contains an equilibrium point.
By the Schur complement, one can confirm that \eqref{IES:cond} and \eqref{integrability} on $D$ implies \eqref{IES:cond2} on $D$.
According to the proof of \cite[Theorem 15]{TRK:18}, if \eqref{IES:cond2} holds on $D$, then there exists $\lambda \in [0, 1)$ such that
\begin{align*}
&\sqrt{(x_{k+1} - x'_{k+1})^\top P (x_{k+1} - x'_{k+1})} \\
&\le \lambda \sqrt{(x_k - x'_k)^\top P (x_k - x'_k)}.
\end{align*}
for all $k=0, 1,\dots$ and $(x_0, x'_0) \in \bR^n \times \bR^n$ as long as $x_k - x'_k \in D$.
This implies that if $D$ is geodesically convex and positively invariant, the closed-loop system is IES on $D$.
Next, if $D$ contains an equilibrium point $x^*$, it follows that
\begin{align*}
\sqrt{(x_1 - x^*)^\top P (x_1 - x^*)} 
\le \lambda \sqrt{(x_0 - x^*)^\top P (x_0 - x^*)}
\end{align*}
for all $x_0 \in D$.
From geodesic convexity, this further implies that $D$ is positively invariant.
Thus, the closed-loop system is IES on $D$.
\red
\end{secrem}

\subsection{Discussions for Generalizations}\label{ss:CD}
In this subsection, we discuss how to generalize our results to the cases where 
the input vector field $b$ is a function of $x$. 
Also, we mention the continuous-time case.

First, we consider the system with non-constant $b$:
\begin{align}\label{sys_b}
x_{k+1} = f (x_k) + b(x_k) u_k,
\quad k \in \bZ_{\ge 0},
\end{align}
where $b: \bR^n \to \bR^n$ is of class $C^1$.
A modification of Proposition~\ref{IES:prop} implies that a controller $u = p(x)$ achieves IES if
there exist $\varepsilon > 0$, $P \in \bR^{n \times n}$,
and $p: \bR^n \to \bR$ of class $C^1$ such that for all $x \in \bR^n$,
\begin{align}\label{IES:cond_b}
&\begin{bmatrix}
P & * \\
( \partial f (x) + p(x) \partial b(x)  + b(x)  \partial p(x) ) P & P
\end{bmatrix}
\succeq \varepsilon I_{2n}.
\end{align} 
The difference from \eqref{IES:cond} is the additional term $p(x) \partial b(x)$.

For finding $P$ first, one can utilize a modification of \eqref{forP}:
\begin{align}\label{IES:cond_b2}
\left\{\begin{array}{l}
P \succ 0 \\
b^+(x) (P - \partial f(x) P \partial^\top f(x)) (b^+(x))^\top \succeq \varepsilon_p I_{n-1}.
\end{array}\right.
\end{align} 
Substituting its solution $P$, $p(x) = m_p (x | Y_p)$,
and $\partial p(x) = \partial m_p (x | Y_p)$ into \eqref{IES:cond_b} yields
an LMI with respect to $Y_p$ and $\varepsilon$ at each $x \in \bR^n$.
Therefore, even for non-constant $b$, one can still design a controller 
only by solving two finite families of LMIs on data points $\{ x^{(i)}\}_{i=1}^N$.

In this paper, we focus on discrete-time systems. 
However, our method can also be applied to the continuous-time systems:
\begin{align}
\dot x = f(x) + b(x) u.
\end{align}
According to \cite[Theorem 1]{FS:14}, 
a controller $u = p(x)$ makes the closed-loop system IES if there exist $\varepsilon > 0$ and $P(x) \succ 0$, $x \in \bR^n$ such that
for all $x \in \bR^n$,
\begin{align}\label{dLyap}
&\sum_{k=1}^n \frac{\partial P(x)}{\partial x_k} (f_k(x) + b_k(x) p(x)) \nonumber\\
&+ P(x) (\partial f (x) + p(x) \partial b(x)  + b(x)  \partial p(x)) \\
&+ (\partial f (x) + p(x) \partial b(x)  + b(x)  \partial p(x))^\top P(x) \preceq -\varepsilon P(x). \nonumber
\end{align}
Let $P_{i,j} \sim \cGP (m_{i,j}(x), k_{i,j}(x,x'))$ and $P_{i,j}^{(l)} = P_{i,j} (x^{(l)}) + \omega_{i,j}^{(l)}$, 
where $P_{i,j}(x)$ denotes the $(i,j)$th element of $P(x)$, and $\omega_{i,j}^{(l)} \sim \cN (0,  \sigma_{i,j}^{(l)})$ is i.i.d.
Denote $P(x|P_{i,j}^{(l)})$ by the posterior mean of $P(x)$ given $\{P_{i,j}^{(l)}\}_{l=1}^N$, $i,j=1,\dots,n$.
Then, we first find $P(x) = P(x|P_{i,j}^{(l)})$ satisfying $P(x) \succ 0$ and
\begin{align*}
&b^+(x) \Biggl( \sum_{k=1}^n \frac{\partial P(x)}{\partial x_k} f_k(x)+ P(x) \partial f (x)  + \partial^\top f (x) P(x) \Biggr)\\
&(b^+(x))^\top \preceq -\varepsilon_p b^+(x) P(x) (b^+(x))^\top,
\quad \forall x \in \bR^n.
\end{align*}
For the obtained $P(x)$, it suffices to solve \eqref{dLyap} with respect to $p(x)= m_p(x|Y_p)$, i.e., $Y_p$.
Therefore, in the continuous-time case, 
nonlinear control design can be achieved only by solving two finite families of LMIs at $\{x^{(l)}\}_{l=1}^N$
even for a non-constant metric $P(x)$.
As mentioned in Section~\ref{PF:sec}, the proposed method can further be applied to various design problems.

\section{Control Design for Unknown Systems}\label{CDUS:sec}
For unknown system dynamics, it is shown by e.g. \cite{FLS:14,FCR:14} that
its state-space model can be estimated from the system's input and output by GPR.
Since GPR is a Bayesian approach, the estimation error is represented by a posterior covariance.
In this section, we show how to compensate for a stochastic learning error by control design.
To focus on exposing the main idea, we consider a case where
the drift vector field is unknown, and the state is measurable, but the results can be generalized to 
the case where all systems dynamics are unknown and 
only the system's input and output are measurable by utilizing the results in \cite{FLS:14,FCR:14}.

\subsection{Learning Drift Vector Fields}
In GPR, we learn each component $f_i$, $i=1,\dots,n$ of $f$ separately from training data $\{x^{(j)}, y_i^{(j)} \}_{j=1}^N$,
\begin{align*}
y_i^{(j)} = f_i(x^{(j)}) +  \omega_{y_i}^{(j)} , 
\quad i=1,\dots,n, \; j=1,\dots,N,
\end{align*}
where $\omega_{y_i}^{(j)} \sim \cN (0, \sigma_{y_i}^2 )$ is i.i.d. 
The number $N$ of training data and training data points $\{x^{(j)}\}_{j=1}^N$ for learning $f$ are allowed to be different from those used for control design. Differently from control design, we can directly obtain training data of $f_i$, and thus it can be estimated by the standard use of GPR. Moreover, to compensate for the learning error of $f_i$ by control design, we estimate the error as the posterior covariance.

We choose a prior distribution of $f_i$ as
$f_i^{\rm pri} \sim \cGP (0, k_i(x,x'))$,
where $k_i:\bR^n \times \bR^n \to \bR$ is a class $C^2$ positive definite kernel. 
Then, the Bayes estimation yields the posterior mean of the joint distribution of $f_i$ and $\partial f_i$ as follows:
\begin{align*}
&\mu_i (x) := \sum_{j=1}^N k_i (x^{(j)}, x) h_i^{(j)} \\
&\partial \mu_i (x) := \sum_{j=1}^N \partial k_i (x^{(j)}, x) h_i^{(j)} \\
&\quad h_i:=( K_i +\sigma_{y_i}^2 I_N)^{-1} Y_i \in \bR^N, \\
&\quad Y_i:=
\begin{bmatrix}
y_i^{(1)} & \cdots & y_i^{(N)}
\end{bmatrix}^\top \in \bR^N \\
&\quad  K_i
:= 
\begin{bmatrix}
k_i(x^{(1)}, x^{(1)})  & \cdots & k_i(x^{(1)}, x^{(N)})\\
\vdots & \ddots &\vdots\\
k_i(x^{(N)}, x^{(1)})  & \cdots & k_i(x^{(N)}, x^{(N)})
\end{bmatrix}
\in \bR^{N \times N},
\end{align*}
where $h_i^{(j)}$ denotes the $j$th component of $h_i$; see, e.g. \cite[Section 2]{RW:06} for the computation of $\mu_i (x)$, 
and $\partial \mu_i (x)$ can be computed by taking its partial derivative with respect to $x$.

\begin{secrem}
When $b$ is also unknown, we learn $g_i(x, u) :=f_i(x) + b_i u$, $i=1,\dots,n$ from training data $\{(x^{(j)}, u^{(j)}), y_i^{(j)}\}_{j=1}^N$,
\begin{align*}
y_i^{(j)} = g_i(x^{(j)}, u^{(j)}) +  \omega_{y_i}^{(j)} , 
\quad i=1,\dots,n, \; j=1,\dots,N.
\end{align*}
To utilize the prior knowledge that $g_i(x, u)$ is linear with respect to $u$, we employ the following kernel $k_i(x, x') + u u'$, where $k_i:\bR^n \times \bR^n \to \bR$ is a class $C^2$ positive definite kernel. 
Namely, we select a prior distribution of $g_i(x, u)$ as $g_i^{\rm pri} \sim \cGP (0, k_i(x, x') + u u')$.
Then, the Bayes estimation yields the posterior mean of $g_i$ as follows:
\begin{align*}
&\bar \mu_i (x, u) := \sum_{j=1}^N (k_i (x^{(j)}, x) + u^{(j)} u) \bar h_i^{(j)} \\
&\quad \bar h_i:=( K_i + K_u +\sigma_{y_i}^2 I_N)^{-1} Y_i \in \bR^N \\
&\quad  K_u
:= 
\begin{bmatrix}
(u^{(1)})^2 & \cdots & u^{(1)} u^{(N)}\\
\vdots & \ddots \\
 u^{(N)} u^{(1)} & \cdots & (u^{(N)})^2
\end{bmatrix}
\in \bR^{N \times N},
\end{align*}
where $K_i$ and $Y_i$ are the same as the above.
Note that $\bar \mu_i(x, u)$ is linear with respect to $u$.
\red
\end{secrem}

A benefit of GPR for learning a function is the ease of analytical computation of the posterior covariance function $v_i:\bR^n \times \bR^n \to \bR$ of $f_i$ as in 
\begin{align}
&v_i (x, x') := k(x, x') - {\bf k}_i^\top (x) ( K_i + \sigma_{y_i}^2 I_N)^{-1}  {\bf k}_i (x') \nonumber\\
&\hspace{4mm} {\bf k}_i (x):=
\begin{bmatrix}
 k_i (x^{(1)}, x) & \cdots & k_i (x^{(N)}, x)
\end{bmatrix}^\top. \nonumber
\end{align}
Similarly, the posterior covariance function $v_{\partial, i}: \bR^n \to \bR^{n \times n}$ of $\partial f_i$ is easy to compute:
\begin{align*}
&v_{\partial, i} (x, x') :=
\partial^2 k_i(x,x') \\
&\hspace{20mm} -
{\bf k}_{\partial, i}^\top (x) (K_i+\sigma_{y_i}^2 I_N)^{-1} {\bf k}_{\partial, i} (x')\\
&\hspace{4mm}
{\bf k}_{\partial, i} (x):= 
\begin{bmatrix}
&\partial_{x'}^\top k_i(x,  x^{(1)}) & \cdots & \partial_{x'}^\top k_i(x, x^{(N)})
\end{bmatrix}^\top.
\end{align*}

Therefore, the posterior distributions of $f_i$ and $\partial f_i$ are respectively obtained by
\begin{align*}
f_i^{\rm post} | Y_i &\sim \cGP ( \mu_i (x), v_i (x, x'))\\
\partial^\top f_i^{\rm post} |Y_i
&\sim \cGP \left(  \partial^\top \mu_i(x), v_{\partial, i} (x, x') \right),
\end{align*}
and consequently,
\begin{align}
f_i^{\rm post}(x) | Y_i &\sim \cN ( \mu_i (x), \sigma_i^2 (x)), \quad \forall x \in \bR^n\nonumber\\
\partial^\top f_i^{\rm post} (x) |Y_i 
&\sim \cN \left(  \partial^\top \mu_i(x), \sigma_{\partial, i}^2 (x) \right), \quad \forall x \in \bR^n \nonumber\\
&\hspace{-14mm}\sigma_i (x) := \sqrt{v_i(x, x)}, \; \sigma_{\partial, i} (x):=\sqrt{v_{\partial, i} (x, x)}. \label{var}
\end{align}
An advantage of obtaining the covariance functions $\sigma_i(x)$ and $\sigma_{\partial, i}(x)$ in nonlinear system identification is that one can compute empirical confidence intervals and decide if one increases training data in some region of interest to relearn the model. Repeating this, one can construct a model with a desired accuracy.

Taking the model learning error into account, a representation of an estimated closed-loop system with $u = p(x)$ becomes
\begin{align}
&x_{k+1} = \mu_c (x_k) + \sigma (x_k) \omega_k\label{gp:sys}\\
&\hspace{4mm}\mu_c (x) := \mu (x) + b p (x) \nonumber\\
&\hspace{5.5mm}\sigma (x) := {\rm diag}\{\sigma_1(x),\dots, \sigma_n(x) \}, \nonumber
\end{align}
where $\omega_k \sim \cN (0, I_n)$ is i.i.d.
The error can be compensated by control design. To see this, we study the stochastic system \eqref{gp:sys} from two aspects. First, by applying a moment IES condition in \cite[Corollary 5.4]{KH:21}, we argue how to choose $\varepsilon > 0$ in the LMI~\eqref{LMI2}. Then, we also discuss how to construct $\bA^{(i)}$, $i =1,\dots, N$ for guaranteeing IES in probability.

\subsection{Moment Incremental Stability}
Proposition~\ref{IES:prop} for IES has been generalized to the moment IES of stochastic systems \cite[Corollary 5.4]{KH:21}. 
This can be used to decide $\varepsilon > 0$ in \eqref{LMI2} for control design. 

\begin{secdefn}\cite[Definition 3.4]{KH:21}
The system~\eqref{gp:sys} is said to be {\it IES in the $p$th moment} if there exist $a>0$ and $\lambda \in (0,1)$ such that
\begin{align*}
\bE [|x_k - x'_k|^p ] \le a \lambda^k |x_0 - x'_0|^p, \; \forall k \in \bZ_{\ge 0}
\end{align*}
for each~$(x_0, x'_0) \in \bR^n \times \bR^n$. 
\red
\end{secdefn}

Applying \cite[Corollary 5.4]{KH:21} to the system \eqref{gp:sys} gives the following condition for moment IES.
\begin{secprop}\label{IES_st:cor}
A system~\eqref{gp:sys} is IES in the second moment if there exist $\bar \varepsilon >0$ and~$\bar P: \bR^n \to \bR^{n \times n}$ such that
\begin{align}
\left\{\begin{array}{l}
\bar P \succ 0 \\
\bar P - \partial^\top \mu_c (x) \bar P \partial \mu_c (x)\\
\hspace{4mm}\succeq \bar \varepsilon I_n + \displaystyle   \sum_{i=1}^n \partial^\top \sigma_i (x) e_i^\top \bar P e_i  \partial \sigma_i (x)
\end{array}\right.
\label{IES_st:cond2}
\end{align}
for all~$x \in \bR^n$, 
where each $e_i$, $i=1,\dots,n$ denotes the standard basis whose $i$th element is $1$, and the other elements are all $0$.
\end{secprop}

\begin{pf}
According to \cite[Corollary 5.4]{KH:21}, the system~\eqref{gp:sys} with i.i.d. noise~$\omega_k$ is IES in the second moment if  there exist $\varepsilon >0$ and~$\bar P: \bR^n \to \bR^{n \times n}$ such that
\begin{align*}
\left\{\begin{array}{l}
\bar P \succ 0 \\
\bE \left[( \partial \mu_c (x) + \partial \sigma (x) \omega_k)^\top \bar P  ( \partial \mu_c (x) + \partial \sigma (x) \omega_k)\right]\\
 - \lambda^2 \bar P \preceq 0
\end{array}\right.
\end{align*}
Since $\omega_k \sim \cN (0, I_n)$ is i.i.d, the left-hand side of the second inequality can be rearranged as
\begin{align*}
&\bE [( \partial \mu_c (x) 
+ \partial \sigma (x) \omega_k)^\top \bar P  ( \partial \mu_c (x) + \partial \sigma (x) \omega_k)] - \lambda^2 \bar P\\
&=\partial \mu_c^\top (x) \bar P \partial \mu_c (x) +
\sum_{i=1}^n \partial^\top \sigma_i (x) e_i^\top \bar P e_i  \partial \sigma_i (x) - \lambda^2 \bar P.
\end{align*}
These inequalities hold if~\eqref{IES_st:cond2} holds.
\qed
\end{pf}

The condition \eqref{IES_st:cond2} can be used to decide $\varepsilon > 0$ in \eqref{IES:cond}, i.e., \eqref{IES:cond2} for control design. 
If \eqref{IES:cond2} holds for a sufficiently large $\varepsilon$ (or $\varepsilon (x)$), then \eqref{IES_st:cond2} holds for some $\bar \varepsilon > 0$. 
Therefore, moment IES suggests how to decide  $\varepsilon > 0$.
 
\subsection{Incremental Stability in Probability}
In Section \ref{ss:CS} for control design, a polytope approach has been mentioned. There is a freedom to design finite families of matrices $\bA^{(j)} \subset \bR^{n\times n}$, $j = 1, \dots, N$. This can be utilized to guarantee IES in probability. 

Applying the Chebyshev's inequality \cite[Theorem 1]{Chen:07} yields, given $c>0$,
\begin{align*}
&\bP \big(\left( \partial f_i^{\rm post}(x) - \partial \mu_i(x) \right) \sigma^{-1}_{\partial, i} (x) \\
&\hspace{4mm} \left( \partial f_i^{\rm post}(x) - \partial \mu_i(x)  \right)^\top < c \big) \ge 1 - \frac{n}{c}\\
&\hspace{30mm}\forall x \in \bR^n, \; i=1,\dots,n,
\end{align*}
where $\sigma_{\partial, i}(x)$ is the variance of the Jacobian matrix, computed in \eqref{var}. Therefore, one can design $\bA^{(j)}$ such that $\partial f_i^{\rm post} (x^{(j)})$ is contained in ${\rm ConvexHull}(\bA^{(j)})$ in probability $(1-n/c)^n$, where note that $\partial f_i^{\rm post} |Y_i$ and $\partial f_j^{\rm post} |Y_j$, $i\neq j$ are mutually independent. For the designed $\bA^{(j)}$, suppose that \eqref{LMI2_cvx} has a set of solutions. Then, a controller guaranteeing IES in probability $(1-n/c)^n$ can be constructed from the solutions.

\section{Example}\label{EX:sec}
\begin{figure}[h]
\begin{center}
\includegraphics[width=80mm]{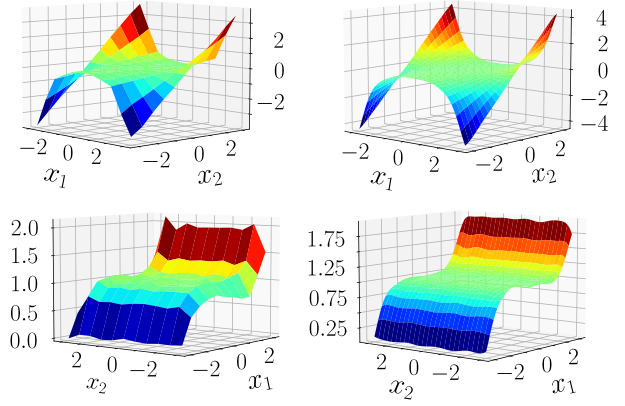}
\caption{(top) Learned $\partial \mu_1/\partial x_1$  (bottom) Learned $\partial \mu_2/\partial x_2$
 (left) $N=121$ (right) $N=2601$}
\label{dmu:fig}
\end{center}
\end{figure}

\begin{figure}[h]
\begin{center}
\includegraphics[width=80mm]{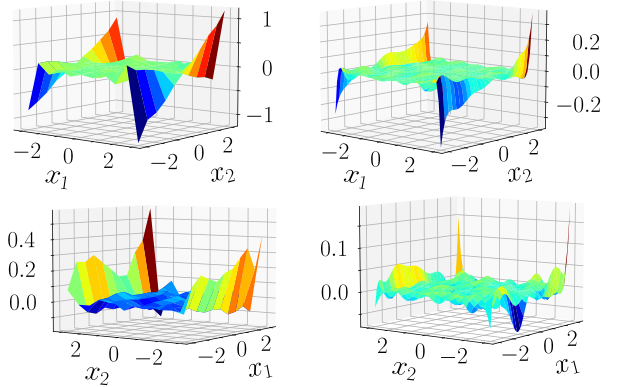}
\caption{(top) $\partial (f_2 -\mu_2)/\partial x_1$   (bottom) $\partial (f_2 -\mu_2)/\partial x_2$ \\
(left) $N=121$ (right) $N=2601$}
\label{err:fig}
\end{center}
\end{figure}

Consider a negative resistance oscillator \cite[Exercise 2.7]{Khalil:96}.
Its forward Euler discretization with the sampling period $\Delta t = 0.01$ is given by
\begin{align*}
f(x) &= x + \begin{bmatrix}
x_2\\
-x_1 + h(x_1) x_2\\
\end{bmatrix} \Delta t, \;
b= \begin{bmatrix}
0 \\  1
\end{bmatrix} \Delta t\\
&h(x_1) = -x_1 + x_1^3 - x_1^5/5 + x_1^7/105.
\end{align*}
We learn $\partial f_2$ only, since $\partial f_1 = [\begin{matrix}0 & 1 \end{matrix}]$ is determined by the psychical structure.
For the number of training data, we consider two cases $N=121$ and $N=2601$.
For both cases, training data points $\{x^{(j)}\}_{j=1}^N$ are equally distributed on $[-3, 3] \times [-3, 3]$. Training data is generated by $y^{(j)}= f_2(x_1^{(j)}, x_2^{(j)}) + \omega_{y^{(j)}}$, where $\omega_{y^{(j)}} \sim \cN (0, 0.01^2)$. As a kernel function, we use a Gaussian kernel $k = e^{-|x-x'|^2/2}$. Figures~\ref{dmu:fig} and~\ref{err:fig} show the learned $\partial \mu_2$ and the learning error $\partial (f_2 - \mu_2)$, respectively. In Fig.~\ref{err:fig}, the error is large around the edges. This is because when computing derivatives at some point, we need information around it, but around the edges, this is not possible. In other words, the learned $\partial \mu_2$ in Fig.~\ref{dmu:fig} is closed to the true $\partial f_2$ except for the edges.

\begin{figure}[tb]
\begin{center}
\includegraphics[width=80mm]{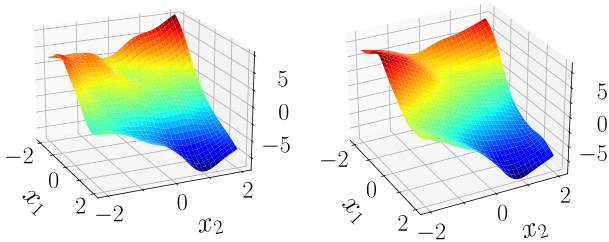}
\caption{Constructed $p$ (left) $N=49$  (right) $N=961$}
        \label{p:fig}
\end{center}
\end{figure}

\begin{figure}[tb]
\begin{center}
        \includegraphics[width=80mm]{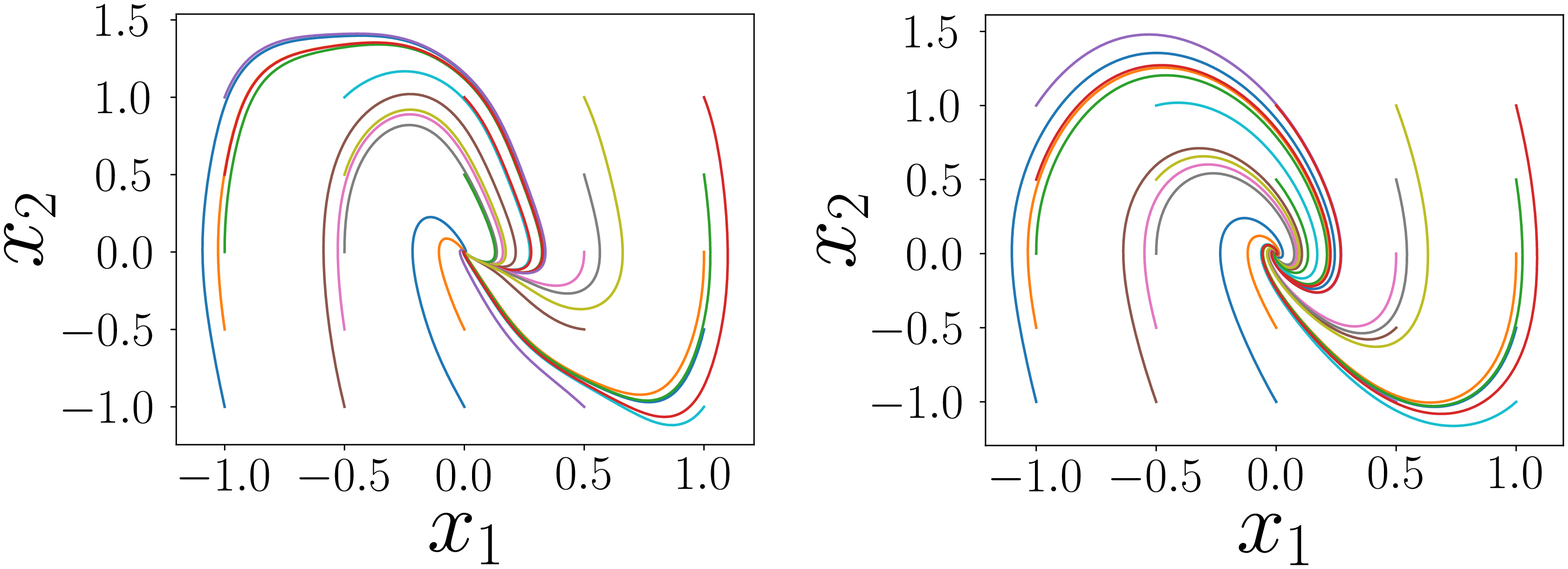}
        \caption{Phase portraits of the closed-loop system (left) $N=121$ for a model  and $N=49$ for a controller  (right) $N=2601$ for a model  and $N=961$ for a controller}
        \label{cl:fig}
\end{center}
\end{figure}

Next, we design a nonlinear controller based on the learned $\partial \mu_2$ by using Algorithm~\ref{alg:main}. To avoid the edges, we consider a smaller region $[-2, 2] \times [-2, 2]$.
For the number of training data, we consider two cases $N=49$ and $N=961$.
For both cases, training data points $\{x^{(j)}\}_{j=1}^N$ are equally distributed. We select $k_0 = e^{-|x-x'|^2/2}$ and $\sigma_p = 0$. Then, for both cases, solutions to the LMI \eqref{LMI1}  are the same:
\begin{align*}
P =  \begin{bmatrix}
30.3 & -25.2\\
-25.2 & 30.0
\end{bmatrix}.
\end{align*}
For this $P$, we solve the LMI \eqref{LMI2} and construct $p(x)$ that is plotted in Fig.~\ref{p:fig}. We apply the constructed controller to the true system. Since the origin is an equilibrium point of $x(k+1) = f(x(k))$, we modify the controller as $u = p(x) - p(0)$ to preserve the equilibrium point. For the different numbers of training data, Fig.~\ref{cl:fig} shows the phase portraits of the closed-loop systems, i.e., the state trajectories starting from different initial states. In each case, the origin of the true system is stabilized by a controller designed for a learned model.

\begin{figure}[tb]
\begin{center}
\includegraphics[width=80mm]{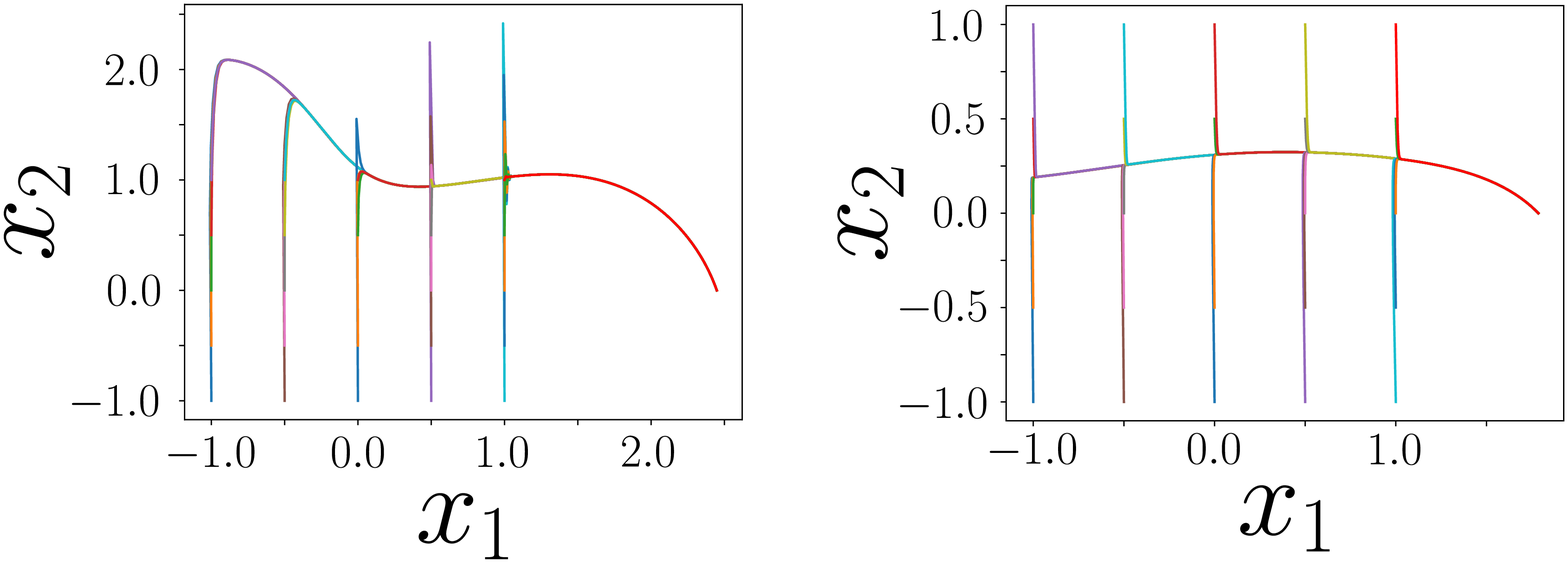}
\caption{Phase portraits of the closed-loop system by $u= - \mu_2 -49.8x_1 + 40.6x_2$ (left) $N=121$ for a model  (right) $N=2601$ for a model}
\label{clfb:fig}
\end{center}
\end{figure}

An advantage of our approach is that a nonlinear stabilizing controller is designed only by solving LMIs. The papers \cite{UH:19,UPH:18} propose stabilizing control design methods for a model learned by GPR. The essence of these methods are to cancel nonlinear terms by state feedback like feedback linearization. We apply this approach. Namely, we divide control design procedure into two steps. First, we apply $u= - \mu_2 + \bar u$ for cancelling the nonlinear term $f_2$, where $\mu_2$ is an estimation of $f_2$. Next, we design linear feedback $\bar u = K x$ based on the linear terms, which can be done by solving an LMI. In fact, we obtain $K=[-49.8 \; 40.6]$. For the different numbers of training data, Fig~\ref{clfb:fig} shows the phase portrait of the closed-loop system by $u= - \mu_2 -49.8x_1 + 40.6x_2$.
In each case, some trajectories (e.g. the red colored one) does not converge to the equilibrium point. 
This can be caused by the gap between true $f_2$ and learned $\mu_2$ as known that feedback linearization is weak at model uncertainty.

\section{Conclusion}\label{Con:sec}
In this paper, we have established an LMI framework for contraction-based control design by utilizing the derivatives of GPR to solve the integrability problem, as an easy-to-implement tool for nonlinear control.
Differently from the standard use, we have used GPR to parametrize a set of controllers and have found suitable parameters from the contraction condition.
The proposed method can further deal with the case where system dynamics are unknown by learning them by GPR, and 
the learning errors can be compensated by control design by simple modifications of the proposed LMI framework.
Future work includes applying our method to a high-dimensional model with a real data set.

\bibliographystyle{plain}
\bibliography{Ref}

\begin{thebibliography}{10}

\bibitem{AJP:16}
V.~Andrieu, B.~Jayawardhana, and L.~Praly.
\newblock Transverse exponential stability and applications.
\newblock {\em IEEE Transactions on Automatic Control}, 61(11):3396--3411,
  2016.

\bibitem{BKH:19}
T.~Beckers, D.~Kuli{\'c}, and S.~Hirche.
\newblock Stable {G}aussian process based tracking control of
  {E}uler--{L}agrange systems.
\newblock {\em Automatica}, 103:390--397, 2019.

\bibitem{Berntorp:21}
K.~Berntorp.
\newblock Online {B}ayesian inference and learning of {G}aussian-process
  state--space models.
\newblock {\em Automatica}, 129:109613, 2021.

\bibitem{Bishop:06}
C.~M. Bishop.
\newblock {\em Pattern Recognition and Machine Learning}.
\newblock Springer, New York, 2006.

\bibitem{BEF:94}
S.~Boyd, L.~El Ghaoui, E.~Feron, and V.~Balakrishnan.
\newblock {\em Linear Matrix Inequalities in System and Control Theory}.
\newblock SIAM, Philadelphia, 1994.

\bibitem{BMT:20}
M.~Buisson-Fenet, V.~Morgenthaler, S.~Trimpe, and F.~Di Meglio.
\newblock Joint state and dynamics estimation with high-gain observers and
  {G}aussian process models.
\newblock {\em IEEE Control Systems Letters}, 5(5):1627--1632, 2020.

\bibitem{Bullo:22}
F.~Bullo.
\newblock {\em Contraction Theory for Dynamical Systems}.
\newblock Kindle Directed Publishing, 2022.

\bibitem{Chen:07}
X.~Chen.
\newblock A new generalization of {C}hebyshev inequality for random vectors.
\newblock {\em arXiv:0707.0805}, 2007.

\bibitem{FS:14}
F.~Forni and R.~Sepulchre.
\newblock A differential {L}yapunov framework for contraction anlaysis.
\newblock {\em IEEE Transactions on Automatic Control}, 59(3):614--628, 2014.

\bibitem{FS:15}
F.~Forni and R.~Sepulchre.
\newblock Differentially positive systems.
\newblock {\em IEEE Transactions on Automatic Control}, 61(2):346--359, 2015.

\bibitem{FS:18}
F.~Forni and R.~Sepulchre.
\newblock Differential dissipativity theory for dominance analysis.
\newblock {\em IEEE Transactions on Automatic Control}, 64(6):2340--2351, 2018.

\bibitem{FCR:14}
R.~Frigola, Y.~Chen, and C.E. Rasmussen.
\newblock Variational {G}aussian process state-space models.
\newblock {\em Advances in Neural Information Processing Systems}, 27, 2014.

\bibitem{FLS:14}
R.~Frigola, F.~Lindsten, T.B. Sch{\"o}n, and C.E. Rasmussen.
\newblock Identification of {G}aussian process state-space models with particle
  stochastic approximation {EM}.
\newblock {\em IFAC Proceedings Volumes}, 47(3):4097--4102, 2014.

\bibitem{FBT:18}
K.~Fujimoto, H.~Beppu, and Y.~Takaki.
\newblock On computation of numerical solutions to {H}amilton-{J}acobi
  inequalities using {G}aussian process regression.
\newblock {\em Proc.2018 American Control Conference}, pages 424--429, 2018.

\bibitem{GAA:21}
M.~Giaccagli, D.~Astolfi, V.~Andrieu, and L.~Marconi.
\newblock Sufficient conditions for global integral action via incremental
  forwarding for input-affine nonlinear systems.
\newblock {\em IEEE Transactions on Automatic Control}, 67(12):6537 -- 6551,
  2022.

\bibitem{GDC:16}
I.~Goodfellow, Y.~Bengio, and A.~Courville.
\newblock {\em Deep Learning}.
\newblock MIT press, Massachusetts, 2016.

\bibitem{IFT:17}
Y.~Ito, K.~Fujimoto, Y.~Tadokoro, and T.~Yoshimura.
\newblock On stabilizing control of {G}aussian processes for unknown nonlinear
  systems.
\newblock {\em IFAC-PapersOnLine}, 50(1):15385--15390, 2017.

\bibitem{IFY:18}
Y.~Ito, K.~Fujimoto, T.~Yoshimura, and Y.~Tadokoro.
\newblock On {G}aussian kernel-based {H}amilton-{J}acobi-{B}ellman equations
  for nonlinear optimal control.
\newblock {\em Proc.2018 American Control Conference}, pages 1835--1840, 2018.

\bibitem{Kawano:21}
Y.~Kawano.
\newblock Controller reduction for nonlinear systems by generalized
  differential balancing.
\newblock {\em IEEE Transactions on Automatic Control}, 67(11):5856--5871,
  2022.

\bibitem{KBC:20}
Y.~Kawano, B.~Besselink, and M.~Cao.
\newblock Contraction analysis of monotone systems via separable functions.
\newblock {\em IEEE Transactions on Automatic Control}, 65(8):3486--3501, 2020.

\bibitem{KC:22}
Y.~Kawano and M.~Cao.
\newblock Contraction analysis of virtually positive systems.
\newblock {\em Systems \& Control Letters}, 168:105358, 2022.

\bibitem{KH:21}
Y.~Kawano and Y.~Hosoe.
\newblock Contraction analysis of discrete-time stochastic systems.
\newblock {\em arXiv:2106.05635}, 2021.

\bibitem{KCS:21}
Y.~Kawano, C.~K. Kosaraju, and J.~M.~A. Scherpen.
\newblock Krasovskii and shifted passivity based control.
\newblock {\em IEEE Transactions on Automatic Control}, 66(10):4926--4932,
  2021.

\bibitem{KO:17}
Y.~Kawano and T.~Ohtsuka.
\newblock Nonlinear eigenvalue approach to differential {R}iccati equations for
  contraction analysis.
\newblock {\em IEEE Transactions on Automatic Control}, 62(10):6497-- 6504,
  2017.

\bibitem{KS:17}
Y.~Kawano and J.~M.~A. Scherpen.
\newblock Model reduction by differential balancing based on nonlinear {H}ankel
  operators.
\newblock {\em IEEE Transactions on Automatic Control}, 62(7):3293--3308, 2017.

\bibitem{Khalil:96}
H.~K. Khalil.
\newblock {\em Nonlinear Systems}.
\newblock Prentice-Hall, New Jersey, 1996.

\bibitem{Krstic:09}
M.~Krstic.
\newblock On using least-squares updates without regressor filtering in
  identification and adaptive control of nonlinear systems.
\newblock {\em Automatica}, 45(3):731--735, 2009.

\bibitem{LS:98}
W.~Lohmiller and J.-J.~E. Slotine.
\newblock On contraction analysis for non-linear systems.
\newblock {\em Automatica}, 34(6):683--696, 1998.

\bibitem{MS:17}
I.~R. Manchester and J.-J.~E. Slotine.
\newblock Control contraction metrics: Convex and intrinsic criteria for
  nonlinear feedback design.
\newblock {\em IEEE Transactions on Automatic Control}, 62(6):3046--3053, 2017.

\bibitem{PZH:21}
M.~Padidar, X.~Zhu, L.~Huang, J.~Gardner, and D.~Bindel.
\newblock Scaling gaussian processes with derivative information using
  variational inference.
\newblock {\em Advances in Neural Information Processing Systems}, 34, 2021.

\bibitem{PDC:14}
G.~Pillonetto, F.~Dinuzzo, T.~Chen, G.~De Nicolao, and L.~Ljung.
\newblock Kernel methods in system identification, machine learning and
  function estimation: A survey.
\newblock {\em Automatica}, 50(3):657--682, 2014.

\bibitem{RW:06}
C.~E. Rasmussen and C.~K.~I. Williams.
\newblock {\em Gaussian Processes for Machine Learning}.
\newblock MIT Press, Massachusetts, 2006.

\bibitem{RSJ:17}
R.~Reyes-B{\'a}ez, A.~J. van~der Schaft, and B.~Jayawardhana.
\newblock Tracking control of fully-actuated port-{H}amiltonian mechanical
  systems via sliding manifolds and contraction analysis.
\newblock {\em Proc. 20th IFAC World Congress}, pages 8256--8261, 2017.

\bibitem{RIK:17}
E.~S. Robert, T.~Iwasaki, and M.~G. Karolos.
\newblock {\em A Unified Algebraic Approach to Linear Control Design}.
\newblock Routledge, London, 2017.

\bibitem{SKS:21}
D.~Sun, M.~J. Khojasteh, S.~Shekhar, and C.~Fan.
\newblock Uncertain-aware safe exploratory planning using {G}aussian process
  and neural control contraction metric.
\newblock {\em Proc. 3rd Annual Conference on Learning for Dynamics and
  Control}, pages 728--741, 2021.

\bibitem{TF:17}
Y.~Takaki and K.~Fujimoto.
\newblock On output feedback controller design for {G}aussian process state
  space models.
\newblock {\em Proc.56th IEEE Conference on Decision and Control}, pages
  3652--3657, 2017.

\bibitem{TRK:18}
D.~N. Tran, B.~S. R{\"u}ffer, and C.~M. Kellett.
\newblock Convergence properties for discrete-time nonlinear systems.
\newblock {\em IEEE Transactions on Automatic Control}, 63(8):3415--3422, 2018.

\bibitem{TC:19}
H.~Tsukamoto and S.-J. Chung.
\newblock Convex optimization-based controller design for stochastic nonlinear
  systems using contraction analysis.
\newblock {\em Proc. 58th IEEE Conference on Decision and Control}, pages
  8196--8203, 2019.

\bibitem{TCS:20}
H.~Tsukamoto, S.-J. Chung, and J.-J.~E. Slotine.
\newblock Neural stochastic contraction metrics for learning-based control and
  estimation.
\newblock {\em IEEE Control Systems Letters}, 5(5):1825--1830, 2020.

\bibitem{TCS:21}
H.~Tsukamoto, S.-J. Chung, and J.-J.~E. Slotine.
\newblock Contraction theory for nonlinear stability analysis and
  learning-based control: A tutorial overview.
\newblock {\em Annual Reviews in Control}, 52:135--169, 2021.

\bibitem{UH:19}
J.~Umlauft and S.~Hirche.
\newblock Feedback linearization based on {G}aussian processes with
  event-triggered online learning.
\newblock {\em IEEE Transactions on Automatic Control}, 65(10):4154--4169,
  2019.

\bibitem{UPH:18}
J.~Umlauft, L.~P{\"o}hler, and S.~Hirche.
\newblock An uncertainty-based control {L}yapunov approach for control-affine
  systems modeled by {G}aussian process.
\newblock {\em IEEE Control Systems Letters}, 2(3):483--488, 2018.

\bibitem{Schaft:13}
A.~J. van~der Schaft.
\newblock On differential passivity.
\newblock {\em IFAC Proceedings Volumes}, 46(23):21--25, 2013.

\end{thebibliography}
\end{document}